\documentclass[preprint,11pt]{elsarticle}

\usepackage{lineno,hyperref}
\usepackage{amsmath}
\usepackage{amssymb}
\usepackage{amsbsy}
\usepackage{amsfonts}
\usepackage{amsopn}
\usepackage{amstext}
\usepackage{graphicx}
\usepackage{amssymb}
\usepackage{amsfonts}
\usepackage{amsmath}
\usepackage{graphicx}
\usepackage[english]{babel}
\usepackage{color}
\usepackage{slashed}
\usepackage{esint}
\usepackage[dvips]{epsfig}
\usepackage[dvips]{graphicx}
\usepackage{float}
\usepackage{units}
\usepackage{textcomp}

\usepackage{hyperref}
\usepackage{slashed}

\modulolinenumbers[5]

\journal{Physics Letters B}

\begin{document}

\begin{frontmatter}

\title{Bipartite-Finsler symmetries}

\author[ufca]{J. E. G. Silva}

\author[pici]{R. V. Maluf}

\author[pici]{C. A. S. Almeida}

\address[ufca]{Universidade Federal do Cariri(UFCA), Av. Tenente Raimundo Rocha, \\ Cidade Universit\'{a}ria, Juazeiro do Norte, Cear\'{a}, CEP 63048-080, Brasil}

\address[pici]{Universidade Federal do Cear\'a (UFC), Departamento de F\'isica, Campus do Pici, Fortaleza - CE, C.P. 6030, 60455-760 - Brazil}


\begin{abstract}
In this work, we study the symmetries of a Lorentz violating bipartite-Finsler spacetime. By using the Finslerian Killing equation for the bipartite-Finsler metric with Lorentzian signature, we analyze how the anisotropy of the bipartite spacetime modifies the local Lorentz symmetries. The symmetries of the pseudo-Finsler metric allow us to obtain the bipartite algebra and interpreted it as a deformed Lorentz algebra. Causality effects driven by the particle modified dispersion relations are investigated upon Minkowski and the Schwarzschild spacetimes. The background bipartite tensor yields to superluminal velocities and modifies the effective gravitational potential.
\end{abstract}


\end{frontmatter}


\section{Introduction}

A complete description of the structure of space-time and fields dynamics near the Planck scale is still a matter of debate and investigation. Some theories suggest that the cornerstones symmetries of the standard model (SM), the CPT and Lorentz symmetries may be violated in the UV regime. In string theories, the vacuum expectation values (VEV) of tensorial fields at some specific vacua might couple to the SM fields driving Lorentz symmetry effects \citep{KS}. The so-called Very Special Relativity (VSR) \citep{vsr} assumes a sub-group of the Lorentz group, the $DSIM(2)$ whereas for the Doubly Special Relativity a larger group leaving invariant both the speed of the light $c$ and the Planck length $l_P$ is assumed \citep{dsr}. Noncommutative geometries \citep{noncommutative}, Horava theories \citep{horava} and Loop Quantum Gravity theory \citep{lqg} also \textbf{contain} Lorentz symmetry breaking effects.

A broad framework to probe CPT-Lorentz symmetry violation is provided by the so-called Standard Model Extension (SME) \citep{cptviolation}. The Lorentz-violating effects were investigated on several different systems involving neutrinos \citep{neutrinos}, photons \citep{photon}, mesons \citep{neutralmesons}. In the gravitational sector, the Lorentz-violating terms might couple to the metric and the connection yielding to an Einstein-Cartan theory \citep{kosteleckygravity}. In the Riemannian limit, assuming a spontaneous symmetry breaking driven by a vector (bumblebee) field \citep{bumblebee} preserves the Bianchi identities and the influence of LV was investigated in cosmology \citep{bumblebeecosmology}.

A conspicuous feature of Lorentz symmetry violation is the modified dispersion relation (MDR) \citep{kosteleckymdr}. The MDR effects can be encompassed into the geometric structure of the space-time by means of the so-called Finsler geometry \citep{Gibbons,kouretsis,girelli,vacaru,pfeifer}. Unlike Riemann geometry, in Finsler geometry the norm of tangent vector $v$ is computed with a general function $F(x,v)$ called Finsler function \citep{rund,chern}. Applications of Riemann-Finsler geometry can be found in anisotropic optics \citep{optics} and graphene \citep{graphene}.

In SME-based Finsler geometry a Riemann-Finsler function is derived from a point-particle Lagrangian with SME coefficients \citep{kosteleckyfinsler1,ColladayMc,russell,marco}. By means of a Wick rotation, the spin-dependent point-particle Lagrangians provide Riemann-Finsler functions whose departure from the usual Riemannian structure depends on the spin particle \citep{kosteleckyfinsler1}. In the photon \citep{marcofinsler1} and in the scalar sector \citep{edwards} new Riemann-Finsler structures were obtained. A CPT-odd fermion has a Finsler structure given by the well-known Randers space \citep{randers}, whereas a chiral CPT-odd fermion point-particle describes a geodesic motion in a novel Finsler space called $b$-space \cite{kosteleckyfinsler1,foster}. Both Randers and $b$-space have Finsler functions (and hence the local Lorentz violation) driven by background vectors. The effects of the background vector field were investigated in cosmology \citep{randerscosmology} and astrophysics \citep{randersastrophysics}.

The Randers and $b$-space are special cases of a general Finsler geometry called bipartite space, where the Finsler function is defined from the Riemannian metric and a symmetric background tensor $s_{\mu\nu}$ \citep{Kosteleckybipartite,euclides}. This bipartite-Finsler structure provides quartic MDR from CPT-even SME interactions \citep{Kosteleckybipartite}. By considering the geometric features of this Finsler spacetime, we can find new symmetries. Although not Lorentz invariant, the Bipartite symmetries reveal what quantities are left invariant by the presence of the bipartite background tensor.

In this work, we seek for the symmetry transformations that preserve the bipartite-Finsler pseudo-metric with Lorentzian signature. Using the Finsler Killing equation, we find the particle transformations leaving unchanged the particle Lagrangian. We define bipartite generators of translations, boosts and rotations. The generators are modified by the inclusion of the bipartite tensor. The corresponding bipartite algebra is defined showing a resemblance with the $\kappa$-algebra \citep{kalgebra}. A similar analysis has been carried out for the Randers \citep{changmdr,changkilling} and Bogoslovsky \citep{bogoslovsky} Finsler spaces. Moreover, the effects of the bipartite tensor on the momentum and energy conservation, as well as the causality analysis is performed. It turns out that the bipartite geometry yields to superluminal signals, an issue exhibited by other Lorentz violating models \cite{stability}. We discuss the modifications of the bipartite Lorentz symmetry breaking on the bound orbit stability and compare to other Finslerian models \citep{Lammerzahl}.

\section{Bipartite-Finsler space}

The Bipartite-Finsler space is an anisotropic spacetime whose interval for a point timelike particle has the form \citep{Kosteleckybipartite,euclides}
\begin{equation}
ds_B =F_{B}(x,u)dt = (\sqrt{g_{\mu\nu}u^{\mu}u^{\nu}}+\sqrt{s_{\mu\nu}u^{\mu}u^{\nu}})dt,
\end{equation}
where $u^{\mu}=\frac{dx^{\mu}}{dt}$ 
is the 4-velocity, $s_{\mu\nu}=s_{\nu\mu}$ is the background Lorentz violating Bipartite tensor and we are assuming the mostly minus metric signature $(+,-,-,-)$. The Finsler function $F_B$ can be written as $F_{B}(x,u)=\alpha(x,u)+\sigma(x,u)$ where $\alpha(x,u):=\sqrt{g_{\mu\nu}u^{\mu}u^{\nu}}$ is the local Lorentz invariant interval and $\sigma(x,u):=\sqrt{s_{\mu\nu}u^{\mu}u^{\nu}}$ is the Lorentz violating term. For $s_{\mu\nu}=b_{\mu}b_\nu$, the Bipartite tensor provides two copies of the Randers space $F_{R}(x,u)=\sqrt{g_{\mu\nu}u^{\mu}u^{\nu}}\pm b_\mu u^{\mu}$. Another example of Bipartite space is the so-called $b$-space, whose Bipartite tensor has the form $s_{\mu\nu}=b^2 g_{\mu\nu} - b_{\mu}b_\nu$ \citep{foster,Kosteleckybipartite}. These both Finsler spacetimes stem from the classical Lagrangian for fermion under chiral CPT-odd Lorentz violating interactions \citep{Kosteleckybipartite}. The indexes for the Bipartite tensor are raised and lowed with the local Lorentz invariant pseudo-metric $g_{\mu\nu}$.

Let us assume the particle action as an integral over the particle worldline \citep{Kosteleckybipartite,euclides}
\begin{equation}
\label{particleaction}
S_B = m\int{(\sqrt{g_{\mu\nu}u^{\mu}u^{\nu}}+\sqrt{s_{\mu\nu}u^{\mu}u^{\nu}})dt}.
\end{equation}
The action (\ref{particleaction}) inherits the Lorentz violating from the spacetime geometry.

Consider the basis $B'=\{e_{0}, e_{1},e_2, e_3 \}$, formed from the eigenvectors of $s_{\mu\nu}$ i.e., $s^{\mu}_{\nu}(e_\rho)^{\nu}=\lambda_\rho \delta_{\nu}^{\mu}(e_\rho)^{\nu}$. Then, $\mathbf{s}=\lambda_0 e^{0}\otimes e^{0} + \sum_{i=1}^{3}\lambda_i e^{i}\otimes e^{i}$, and therefore $\sigma=\sqrt{\lambda_0 + \sum_{i=1}^{3}\lambda_i (v^{i})^2}$.  In order to guarantee the condition $s_{\mu\nu}u^{\mu}u^{\nu}\geq 0$, we assume that $\lambda_0 \geq 0$ and $\lambda_i \geq 0$, a condition similar to the weak energy condition for the stress-energy tensor. For the Randers space, $s^{\mu}_\nu = b^{\mu}b_\nu$ and the background vector $b^{\mu}$ is an eigenvector of $s^{\mu}_\nu$ with eigenvalue $\lambda=b^2$. Thus, the Randers vector $b^\mu$ defines a privileged direction in spacetime. In Randers spacetime the bipartite tensor is idempotent, i.e.,
\begin{equation}
\label{idempotent}
s^{\mu}_{\rho}s^{\rho}_{\nu} = \zeta s^{\mu}_{\nu},
\end{equation}
where $\zeta=b^2=g_{\mu\nu}b^{\mu}b^\nu$. We assume the condition Eq. \eqref{idempotent} is valid for any bipartite space.

We can define an anisotropic bipartite-Finsler metric $g^{B}_{\mu\nu}(x,u)$ such that $ds^{2}_B = g^{B}_{\mu\nu}(x,u)u^{\mu}u^{\nu}$ by \citep{Kosteleckybipartite}
\begin{eqnarray}
\label{bipartitemetric}
g^{B}_{\mu\nu}(x,u) & = & \frac{1}{2}\frac{\partial^{2}F_{B}^{2}}{\partial u^{\mu}\partial u^{\nu}}\nonumber\\
					& = &  \frac{F}{\alpha}g_{\mu\nu} +  \left(\frac{F}{\sigma}s_{\mu\nu}-\alpha\sigma k_{\mu}k_{\nu}\right),
\end{eqnarray}
where, $\alpha:=\sqrt{g_{\mu\nu}u^{\mu}u^{\nu}}$, $\beta:=\sqrt{s_{\mu\nu}u^{\mu}u^{\nu}}$ and $k_{\mu}=\left(\frac{g_{\mu\nu}}{\alpha}-\frac{s_{\mu\nu}}{\sigma}\right)u^{\nu}$. The bipartite-Finsler metric encompasses the bipartite background tensor and its Lorentz violating effects.
The inverse bipartite-Finsler metric which satisfies the condition $g^{B\mu\rho}g^{B}_{\rho\nu}=\delta^{\mu}_{\nu}$ have the form \citep{Kosteleckybipartite}
\begin{eqnarray}
g^{B\mu\nu}(x,u) & = & \frac{\alpha}{F}\Big[g^{\mu\nu}+\left(\frac{\sigma^{\perp}}{\sigma}\right)^2 \frac{\alpha}{S}\lambda^{\mu}\lambda^{\nu}-\frac{\alpha}{S}s^{\mu\nu}\Big],
\end{eqnarray}
where $\lambda_\mu :=(1/\sigma^{\perp})\left(s_{\mu\nu}-(\sigma S/F)g_{\mu\nu}\right)u^\nu$, $\sigma^\perp :=\sqrt{(\zeta g_{\mu\nu}-s_{\mu\nu})u^\mu u^\nu}$ and $S:=\zeta \alpha +\sigma$ \citep{Kosteleckybipartite}.

The directional dependence of the bipartite metric can be measured by the so-called Cartan tensor $C_{\mu\nu\rho}$, defined as \citep{Kosteleckybipartite}
\begin{eqnarray}
C_{\mu\nu\rho} & = & \frac{F_B (x,u)}{2}\frac{\partial g^{B}_{\mu\nu}(x,u)}{\partial u^{\rho}}\nonumber\\
				& = & \frac{1}{\alpha\sigma}g_{\mu\nu}s_{\rho\lambda}u^\lambda - \frac{\sigma}{\alpha^3}g_{\mu\nu}u_\rho+\frac{1}{\alpha\sigma}s_{\mu\nu}u_\rho -\left(\frac{\sigma}{\alpha}u_\rho + \frac{\alpha}{\sigma}s_{\rho\lambda}u^\lambda \right)k_\mu k_\nu +\nonumber\\ 
				& - & \alpha\sigma \left(\frac{\partial k_\mu}{\partial u^\rho}k_\nu + \frac{\partial k_\nu}{\partial u^\rho}k_\mu\right).	
\end{eqnarray}
In fact, the Cartan tensor vanishes for the usual Riemannian geometry.

Since the bipartite metric depends on both the position $x$ and the velocity $u$, a complete description of this Finsler spacetime demands we interpreted it as a configuration space $(x,u)$ and consider both position and direction variation, respectively $\frac{\partial}{\partial x^\mu}$ and $\frac{\partial}{\partial u^{\mu}}$ \cite{vacaru}.
Consider the so-called horizontal derivative
\begin{equation}
\label{horizontalderivative}
\delta_{\mu}:=\frac{\partial}{\partial x^{\mu}} - N_{\mu}^{\nu}\frac{\partial}{\partial u^\nu}
\end{equation}
and the vertical derivative $\bar{\partial
}_{\mu}:=F\frac{\partial}{\partial u^\mu}$, where $N^{\mu}_{\nu}:=\gamma^{\mu}_{\nu\lambda}\frac{u^\lambda}{F}-C^{\mu}_{\nu\lambda}\gamma^{\lambda}_{\delta\epsilon}(u^{\delta}u^{\epsilon}/F^2)$ and the usual Christoffel symbols are $\gamma^{\mu}_{\nu\lambda}:=\frac{g^{B\mu\rho}}{2}(\partial_{\nu}g^{B}_{\rho\lambda}+\partial_\lambda g^{B}_{\rho\nu}-\partial_\rho g^{B}_{\nu\lambda})$ \cite{chern}. The basis $(\delta_\mu , \bar{\partial}_{\mu})$ is orthogonal and splits the configuration space $TM_{4}/0=\{(x,u), x\in M_4,  u\in T_{x}M_4\}$ into a horizontal space spanned by $(\delta_\mu)$ and a vertical space spanned by $(\bar{\partial}_\mu)$ \citep{chern}. Using the dual basis $(dx^\mu, \delta u^\mu)$, where $\delta u^\mu := (1/F)(du^\mu +N^{\mu}_\nu dx^\nu)$, the configuration space has a metric $ds^2= g_{\mu\nu}^B dx^\mu dx^\nu +(g^{B}_{\mu\nu}/F^2)\delta u^\mu \delta u^\nu$ \citep{vacaru}. Although the tangent bundle $TM/0$ contains the whole metric directional-dependence of the spacetime, it is worthwhile to stress that, for a tangent vector $v^\mu \in T_{x}M_4$ only the horizontal metric defines its norm.

In order to study how the spacetime geometry varies from point to point and explore the effects of these changes upon the particle symmetries and dynamics we have to define a connection $\omega^{\mu}_{\nu}=\Gamma^{\mu}_{\nu\rho}dx^\rho + \Gamma'^{\mu}_{\nu\rho}\delta y^\rho$ \citep{chern,vacaru}. The metric changes as $\nabla g= g^{B}_{\mu\nu|\rho}dx^{\mu}\otimes dx^{\nu}\otimes dx^{\rho}+g^{B}_{\mu\nu;\rho}dx^{\mu}\otimes dx^{\nu}\otimes (\delta u^{\rho}/F)$, where $g^{B}_{\mu\nu|\rho}=\delta_\rho g_{\mu\nu}^B -\Gamma_{\rho\mu}^{\lambda}g_{\lambda\nu}^B-\Gamma_{\rho\nu}^{\lambda}g_{\lambda\mu}^B$
and $g^{B}_{\mu\nu;\rho}=\bar{\partial}_{\rho}g_{\mu\nu}^{B} -\Gamma_{\rho\mu}^{'\lambda} g_{\lambda\nu}^{B} -\Gamma_{\rho\nu}^{'\lambda}g_{\lambda\mu}^{B}$. Assuming a metric compatible connection $\nabla g=0$, we obtain the so-called Cartan connection whose horizontal connection has a familiar form 
\begin{equation}
\label{horizontalconnection}
\Gamma^{B\mu}_{\nu\rho}(x,u):=\frac{g^{B\mu\sigma}(x,u)}{2}\Big[\delta_\nu g^{B}_{\sigma\rho}(x,u)+\delta_\rho g^{B}_{\sigma\nu}(x,u)-\delta_{\sigma}g^{B}_{\nu\rho}(x,u)\Big],
\end{equation}
whereas the vertical connection is given by the Cartan tensor $\Gamma_{\rho\nu}^{'\lambda}=C_{\rho\nu}^{'\lambda}$ \cite{chern,pfeifer}. Therefore, the connection $\Gamma^{B\mu}_{\nu\rho}(x,u)$ preserves the Finsler metric $g^{F}_{\mu\nu}(x,u)$ as we move along the Bipartite space. This fact will be used in the next section to find the local isometries (symmetries) of the bipartite spacetime.

\section{Bipartite transformations}

In this section we study the coordinate transformation which preserves the bipartite particle interval (action) Eq. (\ref{particleaction}). 

The action is clearly invariant upon usual observer Lorentz transformations $x'^{\mu}:=(\Lambda)^{\mu}_{\nu}x^{\nu}$. For particle transformations, we consider an actual displacement of the particle along a trajectory preserving the action. To do so, let us consider the Finsler Killing equation which provides the local symmetries of the bipartite Finsler metric $g^{B}_{\mu\nu}(x,u)$.

The general way to find the local symmetries $\xi^{\rho}$ of the action is through the Lie derivative of the Finsler metric, given by \cite{girelli,pfeifer,rund}
\begin{eqnarray}
\label{liederivative}
\mathcal{L}{_\xi}g^F_{\mu\nu}(x,u)&:=&\xi^{\rho}g^{F}_{\mu\nu|\rho}(x,u) + g^{F}_{\mu\rho}(x,u)\xi^{\rho}_{|\nu}+g^{F}_{\nu\rho}(x,u)\xi^{\rho}_{|\mu}+\\
&+& 2C_{\mu\nu\rho}(x,u)\xi^{\rho}_{|\sigma}u^{\sigma}\nonumber,
\end{eqnarray}
where the Finsler covariant derivative is defined by the horizontal connection (\ref{horizontalconnection}).
A Finslerian Killing vector satisfies $\mathcal{L}{_\xi}g^F_{\mu\nu}(x,u)\equiv 0$ and then, the Killing equation turns into
\begin{equation}
\label{finslerkillingequation}
g^{F}_{\mu\rho}(x,u)\xi^{\rho}_{|\nu}+g^{F}_{\nu\rho}(x,u)\xi^{\rho}_{|\mu}+2C_{\mu\nu\rho}(x,u)\xi^{\rho}_{|\sigma}u^{\sigma}=0.
\end{equation}

It is worthwhile to mention that we are interested in the particle symmetries as the particle moves in $M_4$ and not on $TTM$. Hence, the Killing vector $\xi \in T_{x}M_4$, i.e., $\xi=\xi(x)$ has no $u$-dependence. In ref. \citep{changkilling} the authors study the symmetry effects on the Randers spacetime considering variations on the direction as well.

For active particle transformations, let us consider the displacement vector in the direction of the particle motion, i.e., $\xi^{\rho}=u^{\rho}$. The compatible connection (\ref{horizontalconnection}) guarantees that the 4-acceleration is orthogonal to the 4-velocity, i.e., $u_{|\sigma}^{\rho}u^{\sigma}=0$. Thus, the Killing equation turns to
\begin{equation}
\label{particlekillingequation}
\xi_{\mu|\nu}+\xi_{\nu|\mu}=0.
\end{equation}
The Killing equation for the particle symmetries (\ref{particlekillingequation}) is rather similar to the usual Lorentz invariant Killing equation. The difference lies only on the horizontal connection (\ref{horizontalconnection}).

Let us first consider the case where $g_{\mu\nu}=\eta_{\mu\nu}$ and $s_{\mu\nu}$ are constants.
The covariant horizontal derivative turns into $\xi_{\mu|\nu}=\delta_{\nu}\xi_{\mu}$ and a constant 4-vector $\epsilon^\mu$ satisfies the equation \eqref{particlekillingequation}. Thus, we can use the Killing vector $\epsilon^\mu$ to define a translation operation by
\begin{equation}
T(\lambda):=e^{i(\epsilon^{\mu}P^{F}_{\mu})\lambda},
\end{equation}
where $\lambda$ is a real parameter and the translation generator is defined as
\begin{equation}
\label{translationgenerator}
P_{\mu}^{F}:=-i\delta_{\mu}.
\end{equation}
The infinitesimal translation is $\xi^\rho=\epsilon^{\mu}\delta_{\mu}x^{\rho}=\epsilon^\rho$.

In order to define the bipartite transformations and probe the directional symmetry, we define $\omega_{\mu\nu}:=\delta_{\nu}\xi_{\mu}$, which yields Eq. \eqref{particlekillingequation} to
\begin{equation}
\label{skewsymmetricmatrices}
\omega_{\mu\nu}+\omega_{\nu\mu}=0.
\end{equation}
Eq. \eqref{skewsymmetricmatrices} is also satisfied by the Lorentz generators. The velocity dependence can be introduced by
\begin{eqnarray}
\omega^{F\mu}_{\nu} & := & g^{F\mu\rho}(u)\omega_{\rho\nu}\nonumber\\
					& = & \frac{1}{1+\sigma}\Big[\eta^{\rho\nu}+\left(\frac{\sigma^{\perp}}{\sqrt{S}\sigma}\right)^2 \lambda^{\rho}\lambda^{\nu} - \frac{1}{S}s^{\rho\nu}\Big]\omega_{\nu\sigma}
\end{eqnarray}
where $S:=\zeta +\sigma$, $\sigma^{\perp}:=\pm \sqrt{\zeta - s_{\mu\nu}u^{\mu}u^{\nu}}$ and $\lambda^{\rho}:=\frac{1}{\sigma^{\perp}}\left(s^{\rho}_{\epsilon}-\frac{\sigma S}{F}\delta^{\rho}_{\epsilon}\right)u^{\epsilon}$.
Therefore, the bipartite transformations are the result of the usual Lorentz transformations and the action of the background Lorentz-violating tensor $s_{\mu\nu}$ on the 4-velocity $u^{\mu}$. Accordingly, a Lorentz violating particle motion under a bipartite background  tensor can be regarded as a bipartite-symmetric movement.

There is still another important class of symmetry transformations in bipartite spacetime. Indeed, the Finslerian vielbein $E^{Fa}_{\mu}$ allow us to rewrite $g^{B}_{\mu\nu}(x,u)u^{\mu}u^{\nu}=\eta_{ab}u'^{a}u'^{b}$, where $g^{B}_{\mu\nu}(x,u)=E^{Fa}_{\mu}(x,u)E^{Fb}_{\nu}(x,u)\eta_{ab}$ and $u'^{a}=E^{Fa}_{\mu}u^{\mu}$. In the bipartite spacetime, the first-order vielbein has the form
\begin{equation}
E^{Ba}_{\mu}(x,u)=\sqrt{\frac{F}{\alpha}}E^{a}_\mu + \sqrt{\frac{F}{\zeta\sigma}}s^{a}_\mu +\sqrt{\frac{\alpha\sigma}{-k^2}}k^{a}k_\mu,
\end{equation}
where $E^{a}_\mu$ is the local Lorentzian vielbein and $k_\mu$ is assumed as timelike vector. The bipartite vielbein represents a local anisotropic transformation that transforms the coordinate basis $\{\partial_\mu\}$ into a Finslerian 
orthogonal basis $\{e_{a}=E^{\mu}_{a}\partial_\mu \}$. The vielbein changes the 4-velocity $u^\mu$ into the anisotropic vector $u'^{a}=E^{Fa}_{\mu}u^{\mu}$, such that the Lorentz violation is moved from the spacetime (Finsler metric) into the
$u'^{a}$. This is analogous to the light propagation in anisotropic media where the electric and displacement fields are related by the anisotropic polarization tensor, $D^{i}=\epsilon^{i}_{j}E^{j}$. 

Once we found the infinitesimal bipartite transformations, let us study the proper bipartite generators
where
\begin{eqnarray}
J^{F}_{\mu\nu}&:=& x_{\mu}P_{F\nu}-x_{\nu}P_{F\mu}\nonumber\\
			  & = & J_{\mu\nu} + \frac{m}{\sigma}(x_{\mu} s_{\nu\rho} - x_{\nu} s_{\mu\rho})u^{\rho}.	
\end{eqnarray}
In Randers space, the generators take the form $J^{F}_{\mu\nu}=J_{\mu\nu} + m(x_\mu b_\nu - x_\nu b_\mu)$, which is similar to one found in Ref. \citep{changmdr}. Defining the bipartite generators operators
\begin{equation}
J_{\mu\nu}^{B}= -i(x_{\mu}\delta_{\nu}-x_{\nu}\delta_{\mu}),
\end{equation}
the proper bipartite particle transformation takes the form
\begin{equation}
\Lambda_{BP}(\lambda):=e^{i\frac{1}{2}(J^{F}_{\mu\nu}\omega^{\mu\nu})\lambda}.
\end{equation}


\section{Bipartite algebra}

The Finsler algebra is strongly modified by the anisotropy of the
action. In fact, the momentum algebra is given by
\begin{eqnarray}
[P^F_\mu , P^F_\nu]	&	=	&	-[\delta_\mu , \delta_\nu]=2 \delta_{[\mu} N^\rho_{\nu]}\frac{\partial}{\partial u^\rho}\nonumber\\
					&	=	&	 R^{F\sigma}_{\rho\mu\nu}(x,P^F)P^{F\rho}\frac{\partial}{\partial P^{F\sigma}}.
\label{finslerlinearmomentumcommutator}
\end{eqnarray}

As a matter of fact, the anisotropy causes non-inertial effects measures with the Finslerian horizontal curvature $R^{F\sigma}_{\rho\mu\nu}(x,u)$. Indeed, this results in the nonvanishing commutator in Eq. \eqref{finslerlinearmomentumcommutator}. The commutator (\ref{finslerlinearmomentumcommutator}) differs from one found in Ref. \citep{changmdr} due to the definition of the momentum generator in function of the horizontal derivative $\delta_\mu$ in Eq. (\ref{translationgenerator}).

The angular momentum generators are even more sensible to the anisotropic structure. In fact, for a constant $s_{\mu\nu}$,  the  angular momentum generators satisfy
\begin{eqnarray}
\label{finslerangularmomentumanglebra}
[J^F_{\mu\nu} , J^F_{\lambda\rho}]	&	=	&	-g^F_{\mu\lambda}(P^{F})J^F_{\nu\rho}-g^F_{\nu\rho}(P^{F})J^F_{\mu\lambda}\nonumber\\
									&	+	&g^F_{\mu\rho}(P^{F})J^F_{\nu\lambda} + g^F_{\nu\lambda}(P^{F})J^F_{\mu\rho}.
\end{eqnarray}
Note the presence of the Finsler metric $g^F_{\mu\lambda}(P^{F})$ which enhances the momentum dependence when compared to the 
Ref. \citep{changmdr}.

We can rewrite Eq. \eqref{finslerangularmomentumanglebra} as 
\begin{eqnarray}
[J^F_{\mu\nu} , J^F_{\lambda\rho}]&=&
\frac{\alpha}{F}(-\eta_{\mu\lambda}J^F_{\nu\rho}-\eta_{\nu\rho}J^F_{\mu\lambda} +	\eta_{\mu\rho}J^F_{\nu\lambda} + \eta_{\nu\lambda}J^F_{\mu\rho}) -D^F_{\mu\lambda}(P^{F})J^F_{\nu\rho}\nonumber\\
&-&D^F_{\nu\rho}(P^{F})J^F_{\mu\lambda} D^F_{\mu\rho}(P^{F})J^F_{\nu\lambda} + D^F_{\nu\lambda}(P^{F})J^F_{\mu\rho},
\end{eqnarray}
where $D_{\mu\nu}^F:=g_{\mu\nu}^F-\frac{\alpha}{F}\eta_{\mu\nu}$.
Assuming the background vector is still constant, the commutator between $P^F_\mu$ and $J^F_{\nu\lambda}$ takes the form
\begin{eqnarray}
\label{thirdcommutator}
[P^F_\mu, J^F_{\nu\lambda}]	&	=	&	g^F_{\mu\nu}(P^{F})P^F_\lambda - g^{F}_{\mu\lambda}(P^{F})P^F_\nu.
\end{eqnarray}

It is worthwhile to say that the commutation relations in equations \eqref{finslerlinearmomentumcommutator}, \eqref{finslerangularmomentumanglebra} and \eqref{thirdcommutator} have an analogous form of that of the Lorentz invariant transformations, changing only $J^F_{\mu\nu}\rightarrow J^F_{\mu\nu}$ and $\eta_{\mu\nu}\rightarrow g^F_{\mu\nu}(P^{F})$.

The Finsler metric deforms the algebra between momentum, Randers transformations and position. Indeed, for a constant $s_{\mu\nu}$,
\begin{equation}
[P^F_\mu , x_\nu]=-i\left(g^F_{\mu\nu}(x,P^F)+\delta_\mu(g^F_{\nu\rho}(x,P^F)x^\rho\right),
\end{equation}
\begin{equation}
[J^{F}_{\mu\nu} , x_{\lambda} ]=-i\Big\{x_\mu g^F_{\nu\lambda}(x,P^F)-x_\nu g^F_{\mu\lambda}(x,P^F)\Big\}.
\end{equation}
Since for $\zeta=0$, $g^F_{\mu\nu}(x,P^F)=\eta_{\mu\nu}$, the Randers algebra resembles the deformed Heisenberg-Weyl algebra of the $\kappa$-Minkowski spacetime \cite{kalgebra}.

From the bipartite generators, we define the boost generators
\begin{eqnarray}
K_i^F	&	:=	&	J^F_{0i}=K_i + \frac{m}{\sigma}(x_0 s_{i\rho} - x_i s_{0\rho})u^{\rho},
\end{eqnarray}
and the bipartite angular momentum generator
\begin{eqnarray}
J^F_i	&	:=	&	-\frac{1}{2}\epsilon^{jk}_{i}J^F_{jk} =J_i - \zeta m \epsilon^{jk}_i a_j x_k.
\end{eqnarray}
Remarkably, the background vector $a^\mu$ produces not only an expected background boost generator but an angular momentum generator as well.

At the particle rest frame where $g^F_{00}(u)=1$, the boosts and angular momentum generators satisfy
\begin{equation}
[K^F_i , K^F_j]=\epsilon^k_{ij}J^F_k + 2 g^{F}_{0[i}K^F_{j]},
\end{equation}
\begin{equation}
[K^F_i , J^F_j]=\epsilon^k_{ij}K^F_k + 2 g^{F}_{0[i}J^F_{j]},
\end{equation}
\begin{equation}
[J^F_i , J^F_j]=\epsilon^k_{ij}J^F_k.
\end{equation}
For a timelike background vector $a^\mu$, where we can set $g^F_{0i}(u)=0$, the Randers algebra takes a form of an extension of the Lorentz algebra by $K_\mu \rightarrow K^F_\mu$ and $J_{\mu\nu} \rightarrow J^F_{\mu\nu}$. For a spacelike and lightlike $a^\mu$, the presence of the terms $g^F_{0i}$ provides mixtures between boosts and Randers generators.

\section{Symmetry effects}

Once we found the symmetries of bipartite spacetime let us consider the motion of a massive particle in the bipartite space and the effects of this motion on conserved quantities.

The covariant canonical momentum is $P_\mu ^{B}=(m u_\mu + \frac{1}{\sigma}s_{\mu\nu}u^{\nu})$. Using the local Lorentz-invariant metric $g^{\mu\nu}$, the canonical momentum 4-vector is $\hat{P}^{\mu}=g^{\mu\nu}P_\nu ^B=(\delta{^\mu}_\nu +\frac{1}{\sigma}s^{\mu}_\nu )P^{\nu}$. However, $\hat{P}^{\mu}$ is not preserved as the particle moves. Instead, using the bipartite-Finsler metric, the Finslerian momentum $P^{B\mu}:=g^{B\mu\nu}(x,u)P^{B}_\nu = \frac{m}{F}u^{\mu}$ is a covariant constant 4-vector. 
If we parametrize the particle worldline to have a constant Finsler norm, i.e., assuming $F(x,u)=1$, then $P^{B\mu}=P^{\mu}$. 
That is equivalent to define a bipartite-Finsler proper time $d\tau_B= \beta_{B}dt$, where the bipartite factor is $\beta_B := \sqrt{1-v^2}+\sqrt{\lambda_0 + \sum_{i=1}^{3}\lambda_i (v^{i})^2}$.

The Finslerian momentum $P^{B\mu}$ satisfies the Finsler geodesic equation
\begin{eqnarray}
\frac{DP^{B\mu}}{dt}=0	&\Rightarrow & \frac{dP^{B\mu}}{dt}+\Gamma^{B\mu}_{\nu\rho}(x,u)u^{\nu} P^{B\rho}=0.
\end{eqnarray}
Consider the connection coefficients in the first-order of $s_{\mu\nu}$ and up to the quadratic terms in $u$. The equation of motion (EoM) is given by
\begin{equation}
\frac{dP^{\mu}}{dt}+m\frac{\alpha}{\sigma}(u^{\nu}\partial_\nu s^{\mu}_\rho)u^{\rho}-m\frac{\alpha}{2\sigma}(\partial ^{\mu}s_{\nu\rho})u^{\nu}u^\rho=0,
\end{equation}
and in a static background $s_{\mu\nu}$, it is equivalent to the pair of equations
\begin{eqnarray}
\frac{dE}{dt}&=&-m\frac{\alpha}{\sigma}\vec{v}\cdot \nabla \lambda_0,\\
\frac{d\vec{P}}{dt}&=&m\frac{\alpha}{\sigma}\left(\nabla \lambda_0 - \vec{v}\cdot \nabla(\lambda_i)v^{i}+ \nabla(s_{ij})v^{i}v^j\right).
\end{eqnarray}
In the low velocities regime, the EoM looks rather similar to the Newtonian limit whose potential is given by $m\frac{\alpha}{\sigma} \nabla \lambda_0$. However, note the presence of the terms proportional to the 4-velocity, which can yield to unstable motions.

For the point particle 4-momentum $P^{\mu}=m u^{\mu}$, defining the effective metric tensor $\tilde{g}_{\mu\nu}:=g_{\mu\nu}-s_{\mu\nu}$, the modified dispersion relation (MDR), provided by the coupling $F(x,P)=\sqrt{g_{\mu\nu}P^{\mu}P^{\nu}}+ \sqrt{s_{\mu\nu}P^{\mu}P^{\nu}}=m$, yields to a quartic MDR
\begin{equation}
\label{mdr2}
(\tilde{g}_{\mu\nu}P^\mu P^\nu - m^2)^2=4m^2 s_{\mu\nu} P^\mu P^\nu.
\end{equation}
In the Randers space, the Eq. \eqref{mdr2} leads to the quadratic MDR $P^2 + 2 \lambda (b\cdot P) - \lambda^2 (b\cdot P)^2=m^2$. 

\begin{figure}[htb] 
       \begin{minipage}[b]{0.48 \linewidth}
           \includegraphics[width=\linewidth]{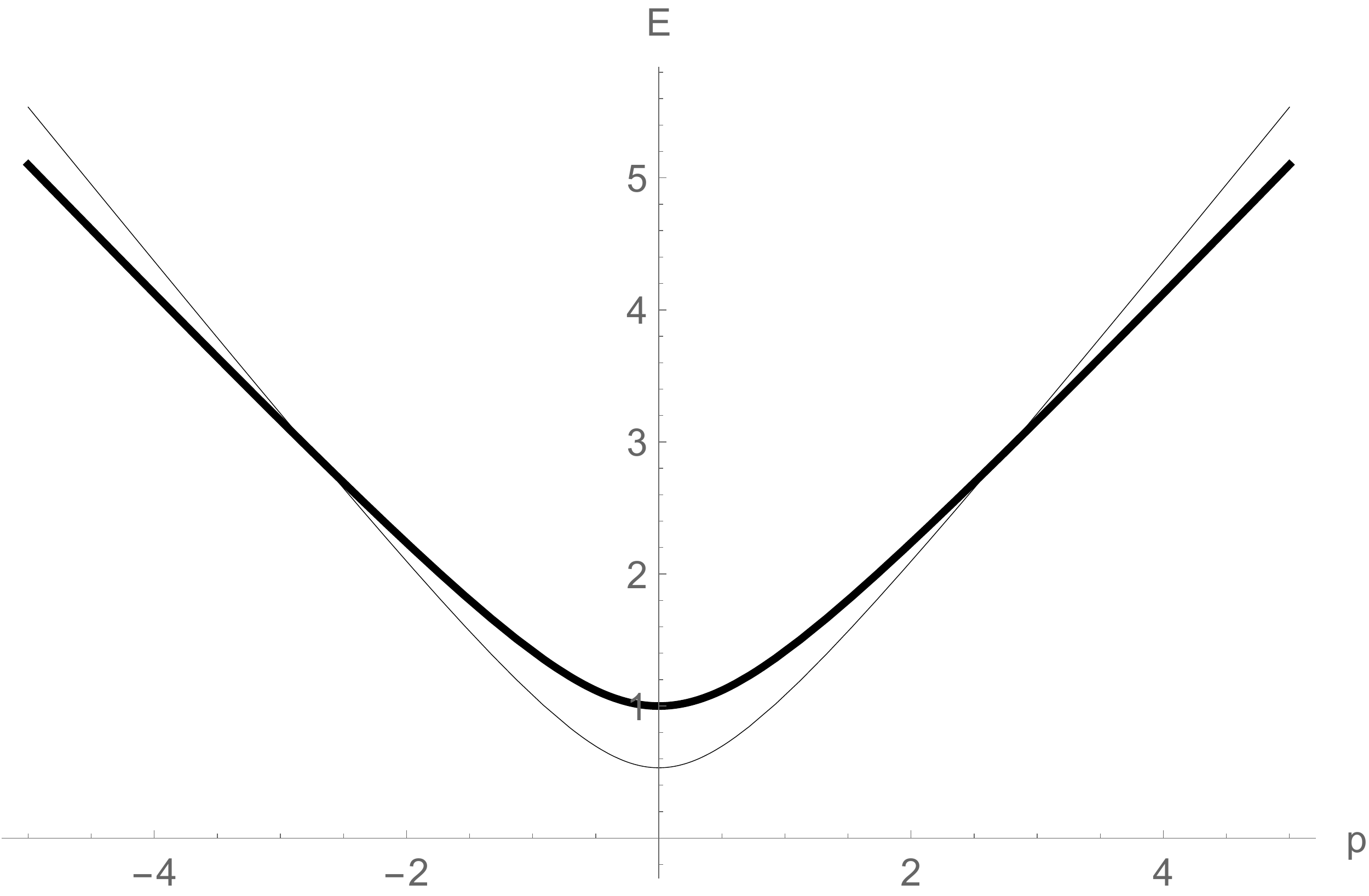}\\
           \caption{Deformed mass-shell for $\lambda_0 =0$ (thick line) and $\lambda_0 =0.4$ (thin line).}
          \label{deformed_mass_shell}
       \end{minipage}\hfill
       \begin{minipage}[b]{0.48 \linewidth}
           \includegraphics[width=\linewidth]{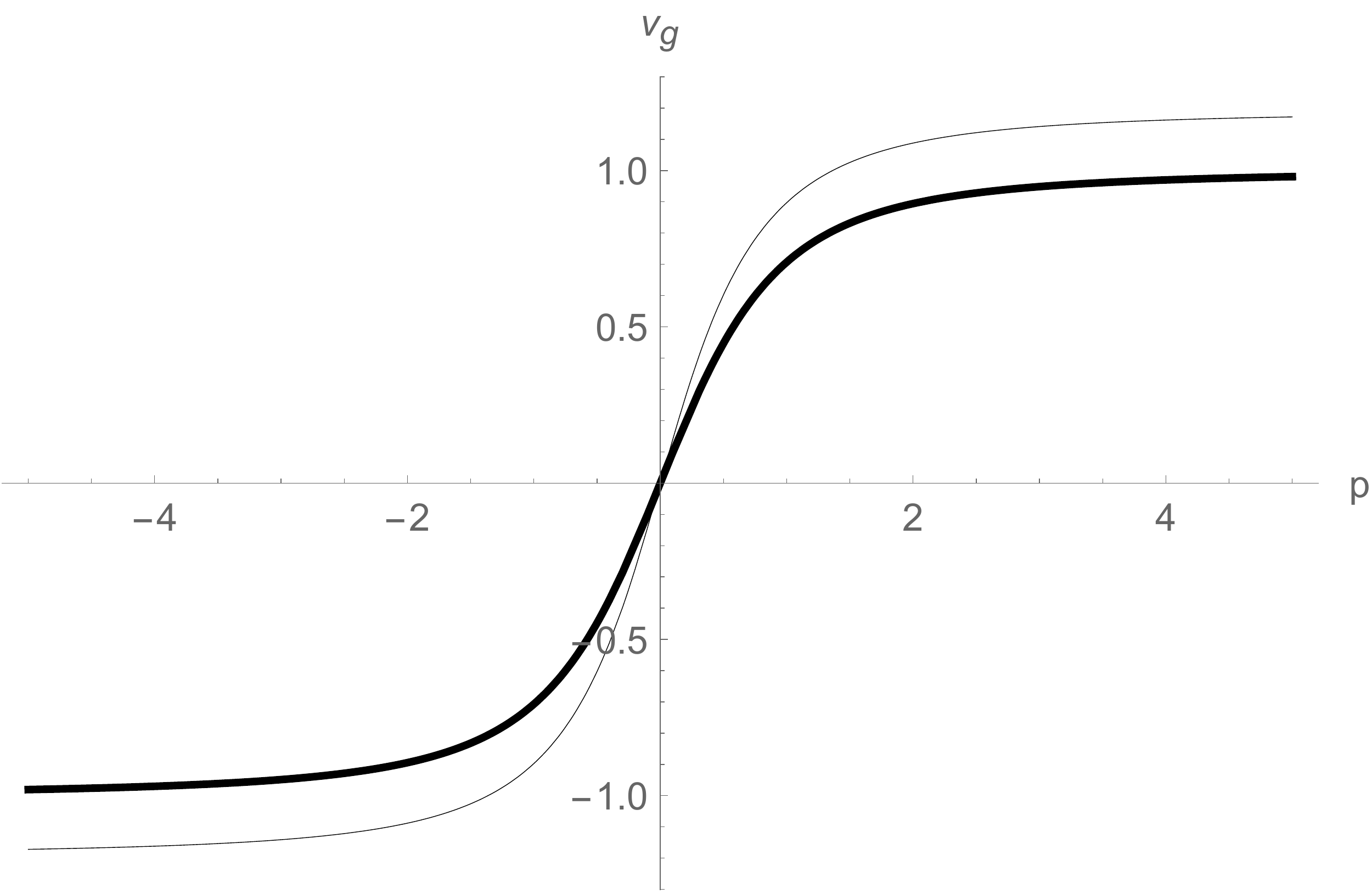}\\
           \caption{Group velocity for $\lambda_=0$ (thick line) and $\lambda_0 =0.4$ (thin line).}
           \label{group_velocity}
       \end{minipage}
   \end{figure}

On the other hand, the modified dispersion relation in a flat Minkowsky space $g_{\mu\nu}=\eta_{\mu\nu}$ \eqref{mdr2} leads to
\begin{equation}
\Big[(1-\lambda_0)E^2 -\sum_{i=1}^{3}(1-\lambda_i) (P^i)^2-m^2\Big]^2 = 4m^2 \Big[\lambda_0 E^2 + \sum_{i=1}^{3} \lambda_i (P^{i})^2\Big].
\end{equation}
The group velocity  $v_{g}^{j} = \frac{dE}{dP_{j}}$ takes the form
\begin{equation}
v^{j}=\Big\{\frac{\delta (1-\lambda_j) + 2m^2 \lambda_j}{\delta (1-\lambda_0) - 2m^2 \lambda_0}\Big\}\frac{P^{j}}{E},
\end{equation}
where $\delta:=(1-\lambda_0)E^2 -\sum_{i=1}^{3}(1-\lambda_i) (P^i)^2-m^2$.
For a pure timelike bipartite tensor $s_{00}=\lambda_0 >0$ and $s_{ij}=0$, 
we plot the deformed mass-shell in figure (\ref{deformed_mass_shell}) and the group velocity in figure (\ref{group_velocity}).

The bipartite geometry exhibits causal issues for high momenta, such as superluminal velocities, a common feature of Lorentz violating models \citep{stability}. 

Now let us consider the effects of the bipartite background tensor upon a particle moving near a massive and spherically symmetric body. Assuming a sufficient weak Lorentz violation such that the geometry is still static and spherically symmetric, the existence of a timelike Killing vector $\zeta^{\mu}=\frac{\partial}{\partial t}$ provides a conserved energy per unit of mass given by $E=g_{\mu\nu}(x)\zeta^{\mu}u^\nu$ whereas the angular symmetry $\psi^{\mu}=\frac{\partial}{\partial\phi}$ yields to a conserved particle angular momentum $L=g_{\mu\nu}\phi^{\mu}\zeta^\nu$. By considering that the Schwarzschild geometry is not altered by the bipartite tensor, the geodesic condition for a timelike particle $g_{\mu\nu}^{B}(x,u)u^{\mu}u^{\nu}=1$ yields to the modified particle dispersion relation
\begin{equation}
\Big[\frac{E^2}{f} - \frac{\dot{r}^2}{f} - \left(1+\frac{L^2}{r^2}+s_{\mu\nu}u^{\mu}u^\nu\right)\Big]^2 =4s_{\mu\nu}u^{\mu}u^\nu,
\end{equation}
where $ds^2 = f(r)dt^2 -f^{-1}dr^2-r^2 d\theta^2 -r^2\sin^{2}\theta d\phi^2$, $u^\mu =\frac{dx^\mu}{d\tau}=(\dot{t},\dot{r},\dot{\theta},\dot{\phi})$ and $f(r)=\left(1-\frac{2GM}{r}\right)$.

In Randers spacetime for  $b^{\mu}=(b^0 , b^r , 0, b^\phi)$ the dispersion relation takes the form
\begin{eqnarray}
\frac{E^2}{2}&=&\frac{\dot{r}^2}{2}+f\Big\{1+\frac{1}{r^2}L^2 +\left(b^0 \frac{E}{f} - f^{-1}b^r \dot{r}-b^\phi L\right)^2 \\\nonumber
&+& (b^0 f^{-1}E - f^{-1}b^r \dot{r} -L b^{\phi})\Big\}.
\end{eqnarray}
The radial component of the Lorentz-violating vector $b^r$ provides velocity-dependent potential term. A background vector with angular component $b^{\mu}=(0,0,0,b^\phi)$ yields to an effective potential whose asymptotic value depends on the test particle angular momentum. For a timelike background vector $b^{\mu}=(b^0,\vec{0})$, the first-order perturbed MDR is
\begin{equation}
E^2=\frac{\dot{r}^2}{2}+f\left(\frac{L^2}{r^2}+1\right)\pm 2b^0 f \sqrt{\dot{r}^2 + f\left(\frac{L^2}{r^2}+1\right)},
\end{equation}
and hence, the effective gravitational potential also exhibits a velocity-dependent term. These examples show that the MDR caused by the bipartite tensor leads to instabilities of particle orbits and asymptotic exotic effects even in the Randers spacetime.

\section{Final remarks and perspectives}

We analyzed the symmetries of the Lorentz violating bipartite-Finsler spacetime. Since the Finsler metric contains corrections due to the background Lorentz violating bipartite tensor, the Finslerian Killing equation enabled us to 
obtain the particle transformations leaving invariant the particle Lagrangian. It is worthwhile to mention that we assumed
particle displacements on the bipartite spacetime $M_4$, not on its tangent bundle $TTM_4$.

We obtained bipartite translations, rotations and boosts, provided the bipartite tensor be constant. The presence of the bipartite tensor in the particle transformations indicates that the Lorentz invariant conserved quantities must be modified by the bipartite tensor in order to be conserved. Moreover, the direction-dependence of the bipartite metric yields to a deformation of the bipartite vielbeins and the Lorentz algebra resembling the $\kappa$-algebra. The quartic dispersion relation yields to causality issues at the UV regime and modifies the gravitational potential by introducing dissipative-like terms. These features suggest the bipartite-Finsler structure might exhibit interesting properties in early cosmology effects.

\section*{Acknowledgments}
\hspace{0.5cm}The authors thank the Conselho Nacional de Desenvolvimento Cient\'{\i}fico e Tecnol\'{o}gico (CNPq), grants n$\textsuperscript{\underline{\scriptsize o}}$ 312356/2017-0 (JEGS), n$\textsuperscript{\underline{\scriptsize o}}$ 305678/2015-9 (RVM) and n$\textsuperscript{\underline{\scriptsize o}}$ 308638/2015-8 (CASA) for financial support.

\end{document}